\PassOptionsToPackage{backref,breaklinks=true,colorlinks,citecolor=blue,urlcolor=[rgb]{0.0,0.0,0.8}}{hyperref}
\documentclass[iop,apjl,appendixfloats,numberedappendix]{emulateapj}
\usepackage{adjustbox}
\usepackage{multirow}
\usepackage{colortbl}
\usepackage[fleqn]{amsmath}
\usepackage{amssymb}
\usepackage{upgreek}
\usepackage{enumerate}
\usepackage{float}
\usepackage{placeins}
\usepackage[USenglish]{babel}
\usepackage{natbib}
\usepackage{graphicx,epsfig}
\usepackage{hyperref}
\usepackage[all]{hypcap}
\DeclareFontFamily{OT1}{pzc}{}
\DeclareFontShape{OT1}{pzc}{m}{it}{<-> s * [1.10] pzcmi7t}{}
\DeclareMathAlphabet{\mathpzc}{OT1}{pzc}{m}{it}

\newcommand{\gl}{\gamma_{\rm L}}
\newcommand{\ec}{\epsilon_{\rm cut}}
\newcommand{\ea}{\epsilon_{\rm A}}
\newcommand{\ed}{\dot{\mathcal{E}}}
\newcommand{\eacc}{E_{\rm acc}}

\newcommand{\pa}{\theta}

\mathchardef\mhyphen="2D

\shorttitle{A Fundamental Plane for Gamma-Ray Pulsars}
\shortauthors{Kalapotharakos et al.}

\begin{document}

\title{A Fundamental Plane for Gamma-Ray Pulsars}

\author{Constantinos Kalapotharakos}
\affil{University of Maryland, College Park (UMCP/CRESST), College
Park, MD 20742, USA} \affil{Astrophysics Science Division,
NASA/Goddard Space Flight Center, Greenbelt, MD 20771, USA}
%\email{constantinos.kalapotharakos@nasa.gov}
\email{ckalapotharakos@gmail.com}

\author{Alice K. Harding}
\affiliation{Astrophysics Science Division, NASA/Goddard Space
Flight Center, Greenbelt, MD 20771, USA}

\author{Demosthenes Kazanas}
\affiliation{Astrophysics Science Division, NASA/Goddard Space
Flight Center, Greenbelt, MD 20771, USA}

\author{Zorawar Wadiasingh}
\affiliation{Universities Space Research Association (USRA)
Columbia, MD 21046, USA} \affiliation{Astrophysics Science Division,
NASA/Goddard Space Flight Center, Greenbelt, MD 20771, USA}

\begin{abstract}
We show that the $\gamma$-ray pulsar observables, i.e., their total
$\gamma$-ray luminosity, $L_{\gamma}$, spectral cut-off energy,
$\ec$, stellar surface magnetic field, $B_{\star}$, and spin-down
power $\ed$, obey a relation of the form
$L_{\gamma}=f(\ec,B_{\star},\ed)$, which represents a 3D plane in
their 4D log-space. Fitting the data of 88 pulsars of the second
\emph{Fermi} pulsar catalog, we show this relation to be
$L_{\gamma}\propto \ec^{1.18\pm 0.24}B_{\star}^{0.17\pm
0.05}\ed^{0.41\pm 0.08}$, a pulsar fundamental plane (FP). We show
that the observed FP is remarkably close to the theoretical relation
$L_{\gamma}\propto \ec^{4/3}B_{\star}^{1/6}\ed^{5/12}$ obtained
assuming that the pulsar $\gamma$-ray emission is due to curvature
radiation by particles accelerated at the pulsar equatorial current
sheet just outside the light cylinder. Interestingly, the FP seems
incompatible with emission by synchrotron radiation. The
corresponding scatter about the FP is $\sim0.35$dex and can only
partly be explained by the observational errors while the rest is
probably due to the variation of the inclination and observer
angles. We predict also that $\ec\propto \ed^{7/16}$ toward low
$\ed$ for both young and millisecond pulsars implying that the
observed death-line of $\gamma$-ray pulsars is due to $\ec$ dropping
below the \emph{Fermi}-band. Our results provide a comprehensive
interpretation of the observations of $\gamma$-ray pulsars, setting
requirement for successful theoretical modeling.
\end{abstract}


\section{Introduction}\label{sec:intro}

Since its launch in 2008, the \emph{Fermi} Gamma-Ray Space
Telescope, has increased by many-fold the number of $\gamma$-ray
pulsars. More specifically, \emph{Fermi} has detected over
230\footnote{https://confluence.slac.stanford.edu/display/GLAMCOG/\\Public+List+of+LAT-Detected+Gamma-Ray+Pulsars}
new $\gamma$-ray pulsars to date \citep[117 of which are included in
the Second Fermi Pulsar Catalog (2PC), ][]{2013ApJS..208...17A}. The
large number of newly discovered $\gamma$-ray pulsars show a number
of trends and correlations among their observed properties, which
probe the underlying physics connected to their emission.

On the theoretical side, there has been tremendous progress in
modeling global pulsar magnetospheres. The Force-Free (FF) solutions
\citep{1999ApJ...511..351C,2006MNRAS.368.1055T,2006ApJ...648L..51S,2009A&A...496..495K}
despite their ideal (i.e., dissipationless) character revealed that
the equatorial-current-sheet (ECS), which emerges at and beyond the
light-cylinder (LC) is a good candidate for the observed
$\gamma$-ray pulsar emission
\citep{2010MNRAS.404..767C,2010ApJ...715.1282B}.

Later studies of dissipative macroscopic solutions
\citep{2012ApJ...749....2K,2012ApJ...746...60L} confirmed, that near
FF-conditions, the ECS is indeed the main dissipative region with
high accelerating electric-field components, $\eacc$. More recently,
the approach of kinetic particle-in-cell (PIC) simulations
(\citealt{2014ApJ...785L..33P}; \citealt{2014ApJ...795L..22C};
\citealt{2016MNRAS.457.2401C}[C16];
\citealt{2018ApJ...855...94P}[PS18];
\citealt{2018ApJ...857...44K}[K18]; \citealt{2018ApJ...858...81B})
confirmed the general picture that $\gamma$-ray pulsars possess a
field structure resembling the FF one while the high-energy emission
takes place near the ECS outside the LC. The advantage of the latter
approach is that it provides particle distributions that are
consistent with the corresponding field structures.

\citet{2014ApJ...793...97K}, \citet{2015ApJ...804...84B}, and
\citet{2017ApJ...842...80K} assuming curvature radiation (CR)
emission from test particles in dissipative macroscopic solutions
were able to reproduce the radio-lag $\delta$ vs. peak-separation
$\Delta$ correlation of the $\gamma$-ray profiles depicted in 2PC
while a comparison between the model and the observed cutoff
energies, $\ec$, revealed a relation between the plasma conductivity
of the broader ECS region as a function of the spin-down power,
$\ed$.

%%%%%%%%%%%%%%%%%%%%%%%%%%%%%%%%%%%%%%%%%%%%%%%%%%%%%%%%%%%%%%%%%%%%%%%%
\begin{figure}[!tbh]
\vspace{0.0in}
  \begin{center}
    \includegraphics[width=1.0\linewidth]{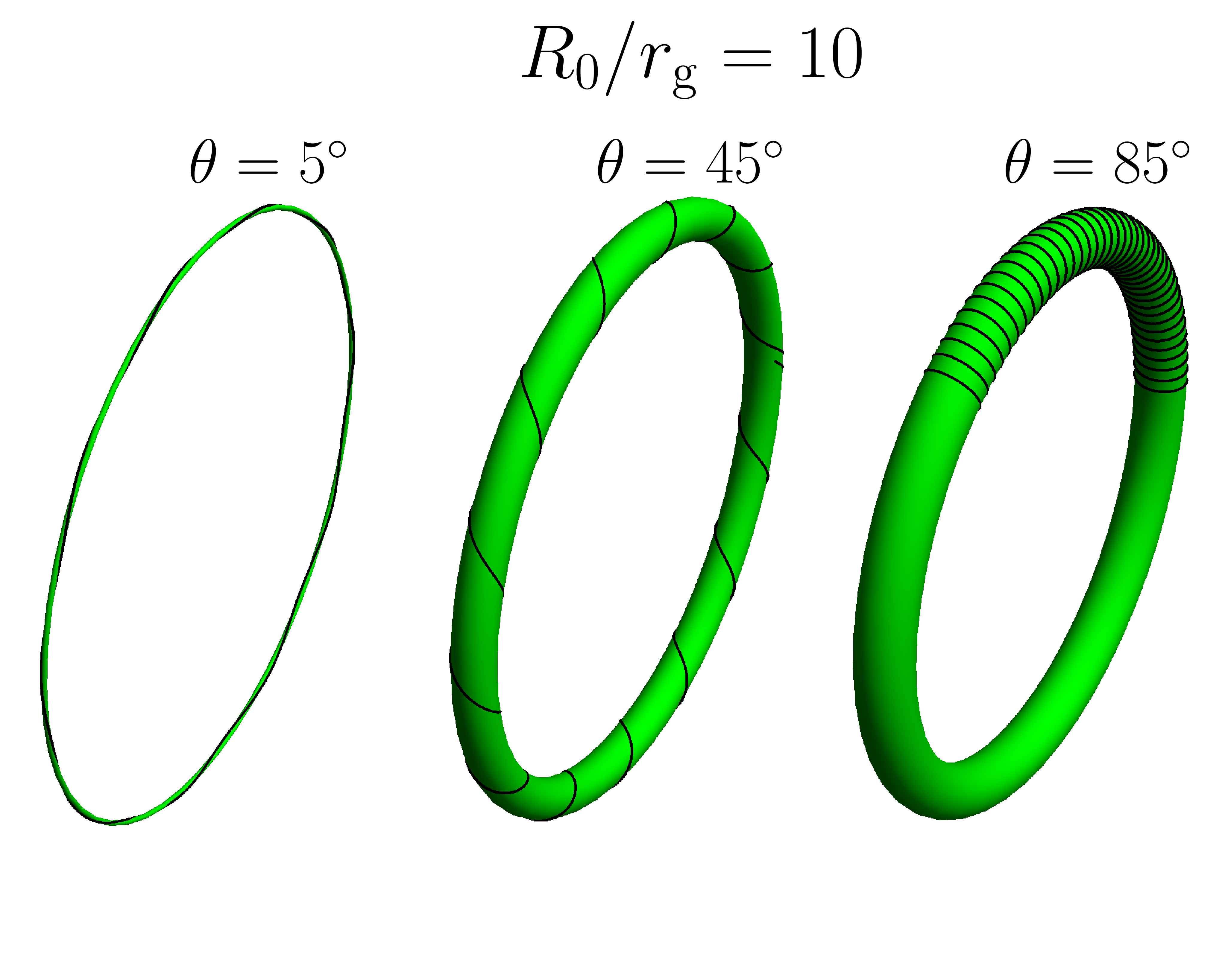}
  \end{center}
  \vspace{-0.5in}
  \caption{The particle orbit for the indicated pitch-angle, $\theta$-values.
  The corresponding motion takes place on a torus with radii $R_{0}$
  and $r_{\rm g}\sin\theta$. For $\theta\to0^{\circ}$, $R_{\rm
  C}\to R_{0}$ while for $\theta\to90^{\circ}$, $R_{\rm
  C}\to r_{\rm g}$.}
  \label{fig01}
  \vspace{-0.0in}
\end{figure}
%%%%%%%%%%%%%%%%%%%%%%%%%%%%%%%%%%%%%%%%%%%%%%%%%%%%%%%%%%%%%%%%%%%%%%%%

The PIC simulations of K18, taking into account the contribution of
CR (by appropriately rescaling the particle energies to realistic
values), revealed a relation between the particle injection rate and
$\ed$ that reproduces the observed range of $\ec$-values (i.e.,
$1-6~\rm{GeV}$).

C16 and PS18 presented PIC simulations of single particle injection
rates and claimed that the corresponding high-energy emission is due
to synchrotron radiation (SR).

Thus, even though there is consensus that the main component of the
observed pulsar $\gamma$-ray emission originates from regions near
the ECS there still is an open question about which radiative
process dominates in the \emph{Fermi} band. Moreover, the recent
detections by \emph{MAGIC} and \emph{HESSII} of very high energy
(VHE) emission from the Crab \citep{2016A&A...585A.133A}, Vela
\citep{2017AIPC.1792d0028D}, and Geminga
\citep{2018_Astrophysics_MAGIC_conference} pulsars imply an
additional emission component, and inverse Compton (IC) seems to be
the most reasonable candidate
\citep{2017ICRC...35..680R,2018ApJ...869L..18H}. In any case, the
multi-TeV photon energies detected imply very high particle energies
($\gl>10^{7}$), which favors CR over SR.

In this letter, we explore the effectiveness of CR and SR to explain
the Fermi spectra, mainly under the assumption that the acceleration
and radiative energy loss occurs in the same location. This is a
different SR-regime from that in C16 and PS18, who assume that
acceleration and radiation, due to reconnection in the ECS, are
spatially uncoupled. Our results show that the observables of all
the \emph{Fermi} pulsars, i.e., young (YP) and millisecond (MP), are
consistent with CR emission. More specifically, our analysis shows
that the \emph{Fermi} YPs and MPs lie on a 3D fundamental plane (FP)
embedded in the 4D space of the total $\gamma$-ray luminosity,
$L_{\gamma}$, $\ec$, the stellar surface magnetic-field,
$B_{\star}$, and $\ed$. This FP is in full agreement with the
theoretical predictions of CR-regime emission.

\section{Reverse Engineering}
\label{sec:rev_eng}

The $\ec$-values observed by \emph{Fermi} provide an excellent model
diagnostic tool. Their variation is small while their value
determination is robust. We note, however, that the $\ec$-values
depend on the adopted spectral fitting model, which in the 2PC reads
$dN/d\epsilon\propto \epsilon^{-\Gamma}\exp(-\epsilon/\ec)$, where
$\Gamma$ is the spectral index. Nonetheless, the apex energies,
$\ea$ of the spectral energy distributions are not much different
than the $\ec$-values corresponding to the model adopted in 2PC.
Actually, $\ea=(2-\Gamma)\ec$ and therefore, only for $\Gamma\approx
2$, $\ea$ deviates considerably from $\ec$. A detailed discussion
about the best fitting function-model goes beyond the scope of this
study. For the rest of the letter, we assume the $\ec$-values
presented in the 2PC, which we believe accurately reflect the
characteristic emission energies.

%%%%%%%%%%%%%%%%%%%%%%%%%%%%%%%%%%%%%%%%%%%%%%%%%%%%%%%%%%%%%%%%%%%%%%%%
\begin{figure*}[!tbh]
\vspace{0.0in}
  \begin{center}
    \includegraphics[width=1.0\linewidth]{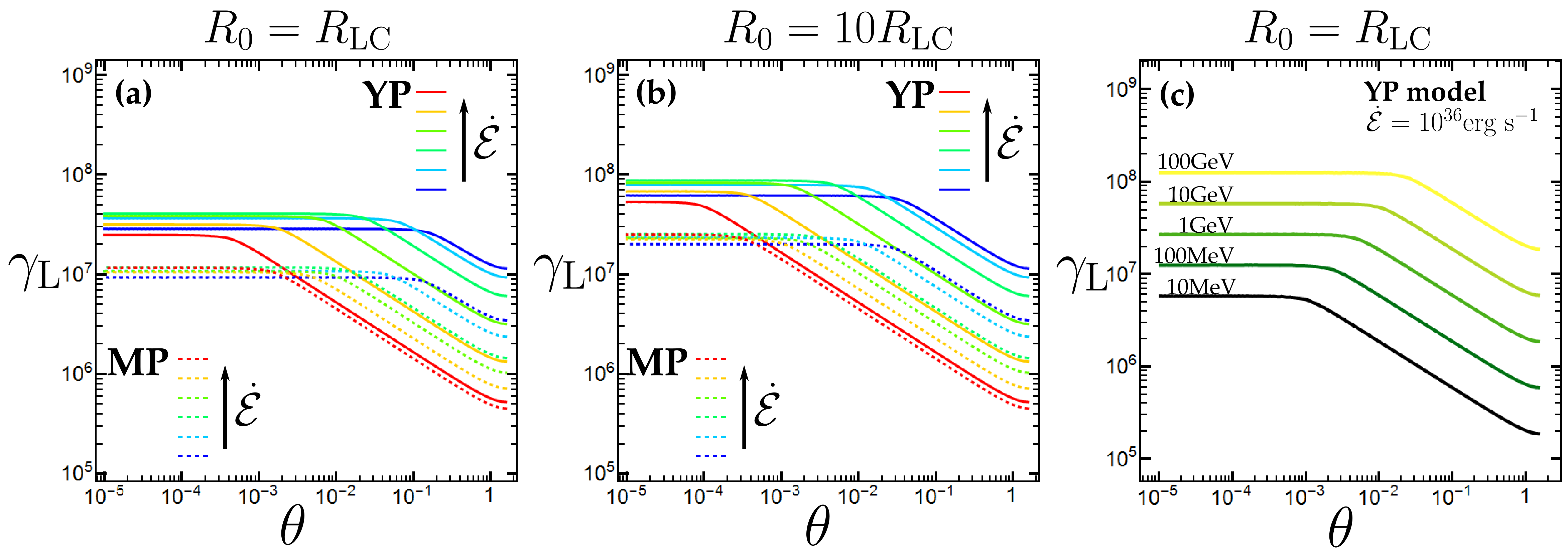}
  \end{center}
  \vspace{-0.25in}
  \caption{\textbf{(a)} The $\gl$ vs. $\theta$ relations that reproduce
  the $\ec$-values corresponding to the different $\ed$-values (different
  colors) for YPs (solid lines) and MPs (dashed lines). These relations
  assume motion at the LC and $R_{0}=R_{\rm LC}$. \textbf{(b)} Similar to
  \textbf{(a)} but for $R_{0}=10R_{\rm LC}$. \textbf{(c)} The $\gl$ vs. $\theta$
  relations for the YP model with $\ed=10^{36}{\rm erg\;s^{-1}}$
  that reproduce the indicated $\ec$-values.}
  \label{fig02}
  \vspace{0.0in}
\end{figure*}
%%%%%%%%%%%%%%%%%%%%%%%%%%%%%%%%%%%%%%%%%%%%%%%%%%%%%%%%%%%%%%%%%%%%%%%%

{We consider a charged particle that is moving in an arbitrary
electromagnetic field. In Appendix~\ref{sec:appendB}, we show that
the trajectory radius of curvature, $R_{\rm C}$, depends mainly on
the maximum field value ($\max(E,B)$) and the generalized
pitch-angle, $\pa$ that measures the deviation of particle velocity
from the locally defined asymptotic trajectory. Below, we assume a
magnetically-dominated field structure where the local $R_{\rm C}$
of the asymptotic flow, which in this case is the guiding-center
trajectory, is $R_{\rm 0}$. The position vector $\mathbf{r}=(x,y,z)$
of a relativistic particle, without loss of generality, can be
locally described by}
\begin{equation}
\label{eq:posvector}
\begin{split}
x&=r_{\rm
g}\sin \pa\;\sin\omega_{\rm g} t\\
y&=(R_{\rm 0}+r_{\rm g}\sin \pa\;\cos\omega_{\rm g}
t)\cos\left(\frac{c}{R_{\rm 0}}\cos\pa\;t\right)\\
z&=(R_{\rm 0}+r_{\rm g}\sin\pa\;\cos\omega_{\rm g}
t)\sin\left(\frac{c}{R_{\rm 0}}\cos\pa\;t\right)
\end{split}
\end{equation}
{with $r_{\rm g}$ the gyro-radius, $\omega_{\rm g}=c/r_{\rm g}$, the
gyro-frequency, and $t$ the time. The motion corresponding to
Eqs.\eqref{eq:posvector} takes place on a 2D torus with radii
$R_{\rm 0}$ and $r_{\rm g}\sin\theta$. Thus, the orbital $R_{\rm C}$
is a function of $\pa$. As $\pa$ goes from 0 to $\pi/2$, $R_{\rm C}$
goes from $R_{\rm 0}$ to $r_{\rm g}$, respectively (see
Fig.\ref{fig01}). We note that particle trajectories corresponding
to different field configurations have similar $(R_{\rm
C},\pa)$-relations taking always into account that the generalized
$r_{\rm g}$ is determined by the corresponding maximum field-value
(Appendix~\ref{sec:appendB}).} The $\ec$-value of the corresponding
spectrum reads
\begin{equation}
\label{eq:ecut} \ec=\frac{3}{2}c\hbar\frac{\gl^3}{R_C(\pa)}
\end{equation}
where $\hbar$ is the reduced Planck constant.

Assuming motion near the LC, we set $R_{\rm 0}=R_{\rm LC}$ and
$B=B_{\rm LC}$. In Fig.\ref{fig02}a, we plot $\gl$ vs. $\pa$, for
different $\ed$-values of YPs and MPs that reproduce the $\ec$
corresponding to the empirical $\ec-\ed$ relations
\begin{equation}
\begin{split}
\label{eq:ec-ed_relations}
\epsilon_{\rm cutYP}&=10^{-103.5+5.75\log\ed-0.0795\log^2\ed}\\
\epsilon_{\rm cutMP}&=10^{-12.47+0.5708\log \ed-0.00571 \log^2\ed}\\
(\epsilon_{\rm cutYP}&,\;\epsilon_{\rm cutMP}\;\text{in GeV and }\ed
\text{ in }\rm{erg\;s^{-1}})
\end{split}
\end{equation}
presented in \citet{2017ApJ...842...80K}\footnote{These expressions
were originally presented with truncated coefficients in fig.2a of
\citet{2017ApJ...842...80K} and therefore, they were not as accurate
as those here.}. Each line corresponds to different combinations of
stellar surface magnetic-field, $B_{\star}$ and period, $P$ (i.e.,
different $\ed$) for YPs (solid lines) and MPs (dashed lines). The
adopted cases (i.e., $B_{\star}, P$ values) are the same as those
presented in Table~2 of K18. More specifically, the $\ed$-values
corresponding to the 6 YP curves are
$$\sim(10^{33},~10^{34},~10^{35},~10^{36},~10^{37},~10^{38})\;\rm
erg\;s^{-1}$$ while those corresponding to the 6 MP curves are
$$\sim(10^{32},~10^{33},~10^{34},~4\times10^{34},~10^{35},~10^{36})\;\rm
erg\;s^{-1}.$$ For each case, a particle should either lie on a
point of these lines or move along these lines in order to emit at
the corresponding $\ec$-value. The $\gl$-value for $\pa\rightarrow
0$ (i.e., CR-regime) does not vary significantly with $\ed$ but is
always higher than the value corresponding to $\pa\rightarrow \pi/2$
(i.e., SR-regime). Moreover, the ratio between the $\gl$-values
corresponding to the two regimes increases with $\ed$.

In Fig.\ref{fig02}b, we show the $\gl-\pa$ relations corresponding
to $R_{\rm 0}=10R_{\rm LC}$. The $\gl$-ratio between the CR and SR
regimes increase by a factor of $\sqrt[3]{10}$. In Fig.\ref{fig02}c,
we plot the $\gl-\pa$ relations for the fourth case of YPs (i.e.,
$\ed\approx 10^{36}\rm erg\;s^{-1}$) that produce the indicated
$\ec$-values. We see that small deviations of $\gl$ and $\pa$ can
significantly change the spectrum $\ec$-value.

%%%%%%%%%%%%%%%%%%%%%%%%%%%%%%%%%%%%%%%%%%%%%%%%%%%%%%%%%%%%%%%%%%%%%%%%
\begin{figure*}[!tbh]
\vspace{0.0in}
  \begin{center}
    \includegraphics[width=1.0\linewidth]{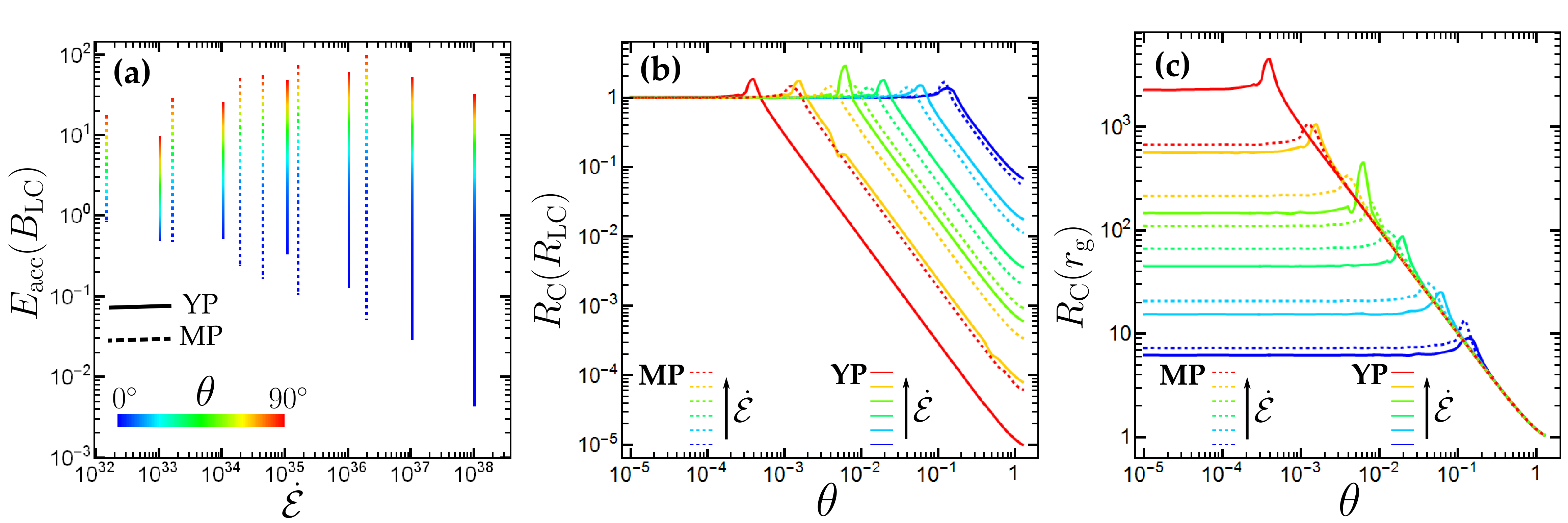}
  \end{center}
  \vspace{-0.25in}
  \caption{\textbf{(a)} The $\eacc$ in the corresponding $B_{\rm LC}$
  units as a function of $\ed$ for MPs (dashed lines) and YPs (solid lines).
  The colors along the lines denote the $\theta$-value according to the
  indicated color-bar. \textbf{(b)} The $R_{\rm C}$ in $R_{\rm LC}$
  units as a function of $\theta$ for the different YP and MP models.
  \textbf{(c)} Similar to \textbf{(b)} but the $R_{\rm C}$ is measured in
  $r_{\rm g}$ units. For all the cases, $R_{0}=R_{\rm LC}$ is assumed.}
  \label{fig03}
  \vspace{0.0in}
\end{figure*}
%%%%%%%%%%%%%%%%%%%%%%%%%%%%%%%%%%%%%%%%%%%%%%%%%%%%%%%%%%%%%%%%%%%%%%%%

In order for particles to continue emitting at the desired $\ec$,
the $\gl-\pa$ constraint should be sustained. In regions of high
acceleration, $\pa$ normally decreases not only because of the
relative {rapid} decrease of the perpendicular momentum component,
which is the result of the radiation-reaction but also because of
the increase of the parallel momentum component, which is the result
of acceleration. The corresponding $\gl$ may increase or decrease
depending on the balance between the radiation-reaction and the
accelerating forces. These variations make the particles divert from
the corresponding $\gl-\pa$ line. {Balancing the radiation losses
with the energy gain due to the accelerating fields,}
\begin{equation}
\label{eq:lossesgain} \frac{2q_{\rm e}^2\gl^4}{3m_{\rm e}c R_{\rm
C}(\pa)^2}=\frac{q_{\rm e}\mathbf{v}\cdot\mathbf{E}}{m_{\rm e}c^2}
\end{equation}
{can preserve $\gl$ {but not $\pa$}. This does not affect the
CR-regime, but for the decreasing segment of the lines
(Fig.\ref{fig02}) the corresponding rapid decrease of $\pa$ (i.e.,
increase of $R_{\rm C}$) tends to destroy the balance and therefore
the $\ec$. Thus, the $\pa$-value should be sustained by another
mechanism (e.g. a heating process). In such a case, the development
of noisy/fluctuating electric components in the perpendicular
direction could in principle sustain $\pa$.}

Taking into account the above assumptions, we can calculate the
$\eacc$ corresponding to each $\pa$-value (assuming preserved
$\gl,~\pa$ values). In Fig.\ref{fig03}a, we plot the $\eacc$ (in
$B_{\rm LC}$ units) for the different YP and MP models (i.e.,
different $\ed$) and for the different $\pa$-values. For small $\pa$
(i.e., CR-regime), $\eacc$ decreases with $\ed$ and it saturates for
smaller $\ed$ to a value $\approx B_{\rm LC}$. For higher $\pa$, the
$\eacc$ increases considerably to a value even above $B_{\rm LC}$.
{In this case, the problem is that the required $\eacc$-value is
well above its upper limit, which is determined by the surrounding
$B$-field (i.e., $B_{\rm LC}$). Nonetheless, for $R_{0}>R_{\rm LC}$
the lower envelope of Fig.~\ref{fig03}a moves towards lower values
allowing larger parts of $\pa>0$ with $\eacc<B_{\rm LC}$.}

In Figs.\ref{fig03}b,c, we plot the $R_{\rm C}$ as a function of
$\pa$ in units of the corresponding $R_{\rm LC}$ and $r_{\rm g}$,
respectively. We see that $R_{\rm C}$ becomes a certain fraction of
$R_{\rm LC}$ ($r_{\rm g}$), for all $\ed$-values, for $\pa\lesssim
10^{-3}$ ($\pa\gtrsim 10^{-1}$). Thus, in the pure CR-regime $R_{\rm
C}\propto R_{\rm LC}$ while in the pure SR-regime $R_{\rm C}\propto
r_{\rm g}$.

\section{The Fundamental Plane of Gamma-Ray Pulsars}
\label{sec:fund_plane}

In Appendix~\ref{sec:append}, we present, for both the CR and SR
processes, relations between $L_{\gamma},~\ec,~B_{\star}$, and
$\ed$, always assuming emission at the LC near the ECS. {These
relations imply the existence of a 3D or 2D (depending on the
regime) FP embedded in the 4D or 3D variable-space.}

The \emph{Fermi}-data allows the investigation of the actual
behavior of the $\gamma$-ray pulsar population. We consider the
function model $L_{\gamma}=A\;\ec^a\; B_{\star}^b\;\ed^d$ and we
calculate the best-fit parameter-values taking into account the 88
2PC YPs and MPs with published $L_{\gamma}$ and $\ec$ values.
Applying the least-squares method in $\log$-space, considering the
same weight for every point, we get the best-fit relation
\begin{equation}
\label{eq:fund_plane_obs} L_{\gamma(3D)}=10^{14.2\pm2.3}\;
\ec^{1.18\pm0.24}\;B_{\star}^{0.17\pm0.05}\;\ed^{0.41\pm0.08}
\end{equation}
where $\ec$ is measured in MeV, $B_{\star}$ in G, and
$L_{\gamma},~\ed$ in $\rm erg\;s^{-1}$. We note that the
$B_{\star}$-values have been derived assuming the FF $\ed$-relation
for the inclination-angle, $\alpha=45^{\circ}$, i.e.,
$B_{\star}=\sqrt{\ed c^3 P^4/4\pi^4 r_{\star}^6 (1+\sin^2
45^{\circ})}$, where $r_{\star}=10^6$cm is the stellar radius. The
best-fit parameters in Eq.\eqref{eq:fund_plane_obs} are extremely
close to those predicted for the CR-regime, $a=4/3,~b=1/6,~d=5/12$
(see Eq.\ref{eq:app08}).

\begin{figure}[!tbh]
\vspace{0.0in}
  \begin{center}
    \includegraphics[width=1.0\linewidth]{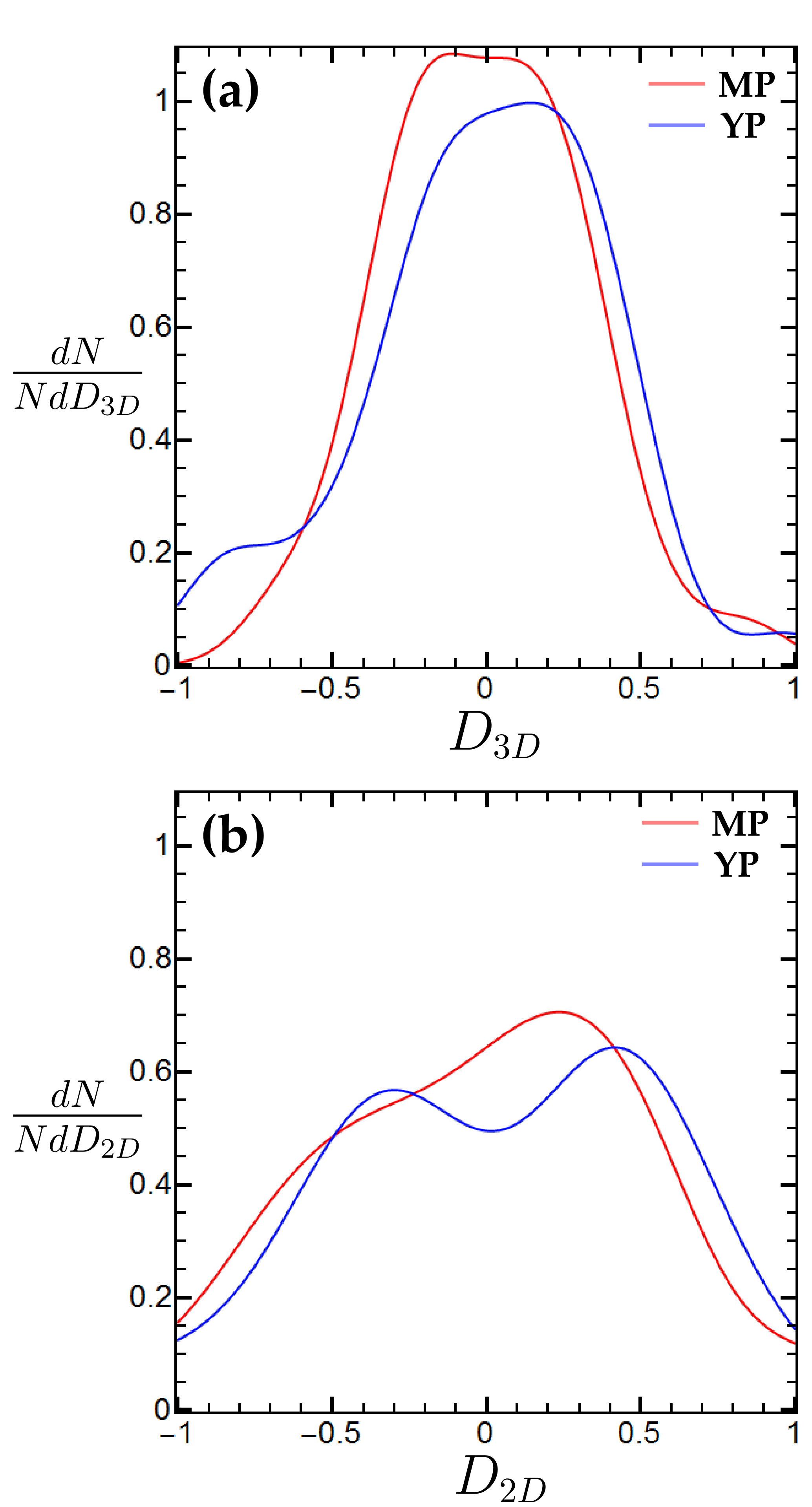}
  \end{center}
  \vspace{-0.1in}
  \caption{\textbf{(a)} Smoothed out distributions of the distances
  $D_{\rm 3D}$ from
  the 3D-FP (Eq.\ref{eq:fund_plane_obs}) for the 2PC MPs (red color)
  and YPs (blue color). \textbf{(b)} Similar to \textbf{(a)} but for
  the 2D-FP (Eq.\ref{eq:fund_plane_obs_3D}).}
  \label{fig04}
  \vspace{0.0in}
\end{figure}

The FP described by Eq.\eqref{eq:fund_plane_obs} applies to the
entire population of $\gamma$-ray pulsars (i.e., YPs and MPs).
Moreover, since the 3D-FP, described by
Eq.\eqref{eq:fund_plane_obs}, is embedded inside a 4D space, it
cannot be easily visualized. In Fig.\ref{fig04}a, we show the
distributions of the signed distances of the observed objects from
this FP for YPs and MPs. The scattering around the FP is similar for
the two classes with a standard deviation of $\sim 0.35 \rm dex$.

The theoretical approach presented in Appendix~\ref{sec:append}
clearly suggests that the dimension of the FP is 3 since it involves
4 variables. Nonetheless, even though our data analysis, which was
motivated by the theoretical findings, resulted in
relation~\eqref{eq:fund_plane_obs}, this doesn't necessary mean that
the effective dimensionality of the data is 3 (i.e., that all the
four variables are necessary to explain the observed data
variation). A quick look at the values of the different variables
makes clear that the range of $\ec$ is intrinsically much smaller
than that of the other variables. Thus, a question that arises is
whether the consideration of $\ec$ provides a better interpretation
of the data-variation.

%%%%%%%%%%%%%%%%%%%%%%%%%%%%%%%%%%%%%%%%%%%%%%%%%%%%%%%%%%%%%%%%%%%%%%%%
%%%%%%%%%%%%%%%%%%%%%%%%%%%%%%%%%%%%%%%%%%%%%%%%%%%%%%%%%%%%%%%%%%%%%%%%

Taking into account the above, we considered a relation
$L_{\gamma}=A\;B_{\star}^b\;\ed^d$ that excludes $\ec$. Then, the
best-fit relation becomes
\begin{equation}
\label{eq:fund_plane_obs_3D}
L_{\gamma(2D)}=10^{15.0\pm2.6}\;B_{\star}^{0.11\pm0.05}\;\ed^{0.51\pm0.09}~.
\end{equation}
In order to compare the two models, we use the Akaike information
criterion \citep[AIC;][]{1974ITAC...19..716A} and the Bayesian
information criterion \citep[BIC;][]{1978AnSta...6..461S}. Both AIC
and BIC measure the goodness of the fit while they penalize the
addition of extra model parameters. The lower the values of AIC and
BIC the more preferable the model is. For the adopted models, the
corresponding AIC, BIC values read
\begin{equation}
\begin{split}
\label{eq:aic_bic_values}
\rm AIC_{3D}&=159,\;\;\;\;\rm AIC_{2D}=180\\
\rm BIC_{3D}&=172,\;\;\;\;\rm BIC_{2D}=189~
\end{split}
\end{equation}
which indicate that the 3D model (i.e., the one that includes $\ec$)
is strongly preferred over the 2D one although the 3D model has an
additional parameter. We note that it is the difference in AIC and
BIC values between the two models that is important rather than
their actual values. The specific AIC difference implies that the
observed sample of data is $e^{(159-180)/2}=e^{-21/2}\approx
10^{-5}$ times less probable to have been produced by the 2D model
than the 3D one. For the BIC any difference greater than ten
indicates a very strong evidence in favor of the model with the
lower value.

In Fig.\ref{fig04}b, we plot similarly to what we did for the
3D-plane, the distributions of the distances of the sample-points
from the 2D-plane \eqref{eq:fund_plane_obs_3D}. We see that these
distributions are not only broader than those of the 3D-model but
they also deviate considerably from the Gaussian shape. We note that
a relation $L_{\gamma}=A\ed^d$ provides results similar to those of
relation~\eqref{eq:fund_plane_obs_3D}.

The last approach provides an unbiased treatment in the sense that
it is data-oriented and dissociated from any theoretical
assumptions. Therefore, the FP, described by
Eq.\eqref{eq:fund_plane_obs}, is supported by the data and could
have, in principle, been discovered without the theory guidance.
Nonetheless, the almost perfect agreement with the theoretical FP,
described by Eq.\eqref{eq:app08} corresponding to the CR-regime,
provides a solid description in simple terms of the physical
processes that are responsible for the phenomenology of $\gamma$-ray
pulsars.

\begin{figure}[!tbh]
\vspace{0.0in}
  \begin{center}
    \includegraphics[width=1.0\linewidth]{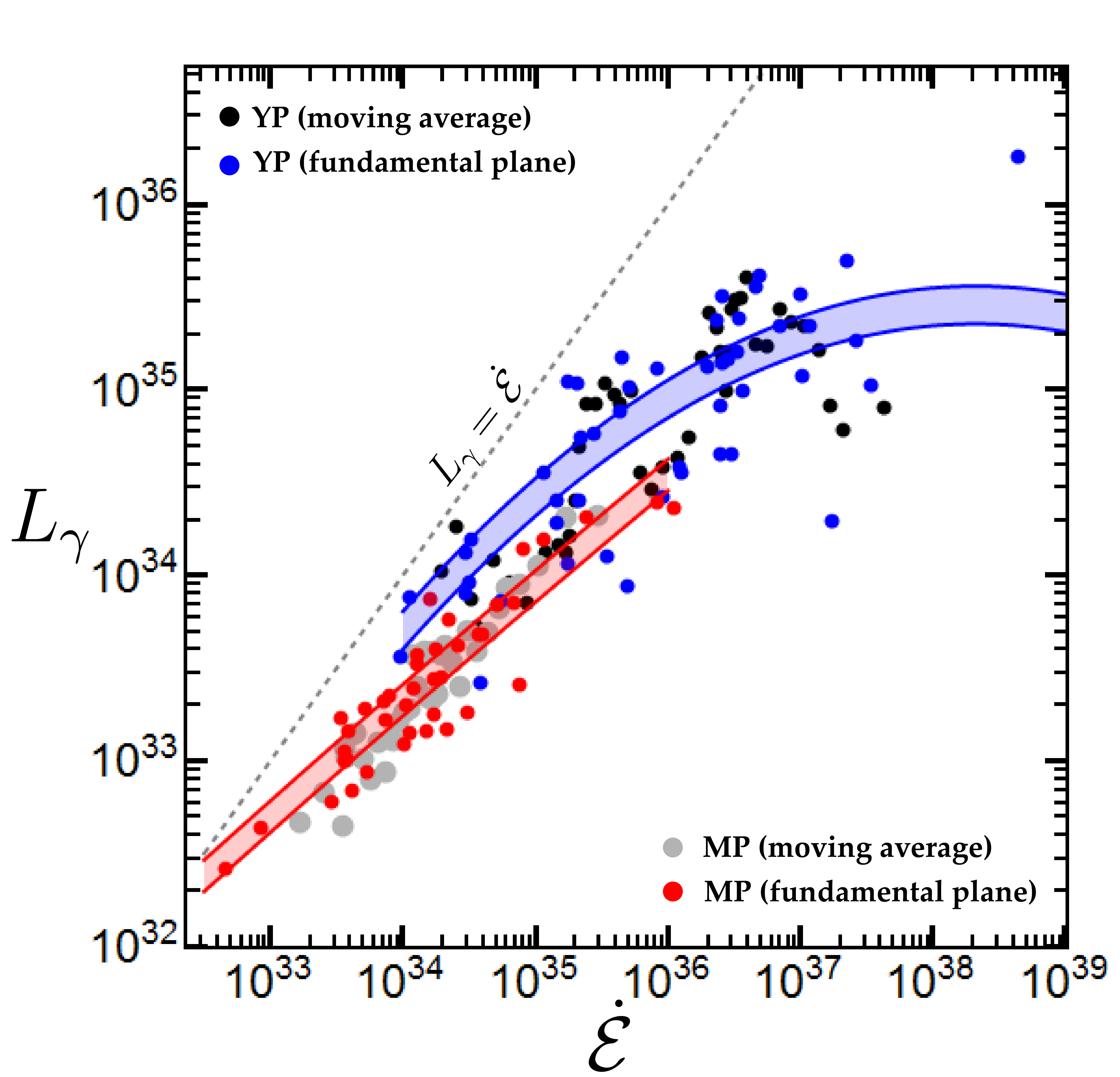}
  \end{center}
  \vspace{-0.1in}
  \caption{The $L_{\gamma}$ vs $\ed$ diagram. The black and gray points
  denote the moving-average values of 2PC YPs and MPs, respectively. The
  blue (2PC-YPs) and red (2PC-MPs) points denote $L_{\gamma}$ values that
  have been calculated by the FP-relation (Eq.\ref{eq:fund_plane_obs})
  taking into account the observed $\ed$, $\ec$, and $B_{\star}$ values.
  The blue (YPs) and red (MPs) zones map the FP-relation
  (Eq.\ref{eq:fund_plane_obs}) assuming that $\ec$ is given by
  Eqs.\eqref{eq:ec-ed_relations} and $B_{\star}$ ranges $10^{8}-10^9$G (for MPs)
  and $10^{11.8}-10^{13}$G (for YPs).}
  \label{fig05}
  \vspace{0.0in}
\end{figure}

In Fig.\ref{fig05}, we reproduce the $L_{\gamma}$ vs. $\ed$ diagram
by calculating the $L_{\gamma}$-values from the FP-relation
\eqref{eq:fund_plane_obs}. Thus, the red and blue points correspond
to the YPs and MPs, respectively, and have been derived using the
corresponding (observed) $B_{\star}$, $\ed$, and $\ec$ values. The
black and gray points show the moving average values (five points
along $\ed$) of 2PC for YPs and MPs, respectively. Finally, the blue
(YPs) and red (MPs) lines have been derived assuming the empirical
$\ec-\ed$ relations \eqref{eq:ec-ed_relations}. The two lines (of
the same color) and the shaded region between them cover the range
of the different $B_{\star}$-values (i.e.,
$B_{\star}=10^{8}-10^{9}$G for MPs and
$B_{\star}=10^{11.8}-10^{13}$G for YPs). We see that the FP-relation
reproduces the observed behavior of $L_{\gamma}$ very well.
Actually, it reproduces the trend of YPs having (on average)
slightly higher $L_{\gamma}$-values than those of MPs for the same
$\ed$ as well as the softening of the $L_{\gamma}$ vs. $\ed$ at high
$\ed$ for the YPs.

Finally, our results indicate that for the CR-regime $\eacc/B_{\rm
LC}$ saturates towards low $\ed$-values (see Fig.\ref{fig03}a and
fig.2b in \citealt{2017ApJ...842...80K}). Assuming that this trend
persists for lower $\ed$, from Eqs.\eqref{eq:ecut} and
\eqref{eq:app04} and taking into account the Eqs.\eqref{eq:app01},
\eqref{eq:app02} for the CR-regime, we get
\begin{equation}
\label{eq:ecut_pred_low_edot} \ec\propto B_{\star}^{-1/8}\ed^{7/16}
\end{equation}
which is a generalization of the eq.(A7) of
\cite{2017ApJ...842...80K}. Taking into account the weak dependence
on $B_{\star}$ and that $B_{\star}$ can be considered more or less
constant for each population (YP or MP), we get
$\ec\propto\ed^{7/16}$, which is not much different than the
empirical behaviors (for low-$\ed$) reflected in the expressions in
Eq.\eqref{eq:ec-ed_relations}. The implied decrease of $\ec$ towards
smaller $\ed$-values where \emph{Fermi} becomes less sensitive
combined with the correspondingly smaller $L_{\gamma}$ provide a
viable interpretation of the (to-date) observed $\gamma$-ray pulsar
death-line \citep[see][]{2019ApJ...871...78S}. Equation
\eqref{eq:app08} (for the CR-regime) and
Eq.\eqref{eq:ecut_pred_low_edot} provide the asymptotic behavior
$L_{\gamma}\propto\ed$, toward low-$\ed$. These claims could be
tested and further explored with a telescope with better sensitivity
in the MeV-band like AMEGO.

%%%%%%%%%%%%%%%%%%%%%%%%%%%%%%%%%%%%%%%%%%%%%%%%%%%%%%%%%%%%%%%%%%%%%%%%
%%%%%%%%%%%%%%%%%%%%%%%%%%%%%%%%%%%%%%%%%%%%%%%%%%%%%%%%%%%%%%%%%%%%%%%%

\section{Discussion and Conclusions}

In this letter, we explore the behavior of particle orbits, for the
entire spectrum of regimes from the pure CR to the pure SR one,
which are consistent with the observed photon energies, adopting the
current consensus that the $\gamma$-rays are produced near the ECS.
The particle $\gl$-values in the CR-regime reach up to $10^7-10^8$
while in the SR-regime and especially for the high $\ed$-values are
2-3 orders of magnitude lower.

Kinetic PIC models also agree with this picture. K18 demonstrated
that in PIC global models, CR emission is produced by particles with
realistic $\gl$-values that reach up to these levels (i.e.,
$10^7-10^8$). Moreover, PS18 claimed that particle emission at GeV
energies is due to SR. Nonetheless, in PS18, the potential drops and
the {corresponding $\eacc$} as are reflected in the presented proton
energies (see fig.6 of PS18)\footnote{{In that study, the protons
are defined as $e^+$, which do not experience radiation-reaction
forces.}} {are (scaled to the actual pulsar environment values)
sufficient to support the $e^{+}, e^{-}$ energies required for the
CR-regime.}

We have derived fundamental relations between $L_{\gamma}$, $\ec$,
$B_{\star}$, and $\ed$ for the pure CR and SR {assuming emission
near the LC at the radiation-reaction regime}. Remarkably, the
\emph{Fermi}-data reveal that the entire pulsar population (YPs and
MPs) lie on a FP that is totally consistent with emission in the
CR-regime. {On the other hand, SR seems to fail at least under the
assumed considerations. Even though SR may work under different
conditions (e.g., acceleration and cooling may occur at different
places), it seems that in such a case, a fine-tuning is needed to
lock not only $\eacc$, the acceleration lengths, the $B$-values, and
the corresponding $\pa$-values where the cooling takes place but
also their dependence on $\ed$ that reproduces the observed
correlations.}

The decrease of the accelerating electric fields (in $B_{\rm LC}$
units) with $\ed$ implies an increasing number of particles that
more efficiently short-out $\eacc$. However, our analysis shows that
for CR the best agreement with observations is achieved when the
number of emitting particles is scaled with the Goldreich-Julian
number-density, $n_{\rm GJ\star}$. Apparently, based on our
considerations in Appendix~\ref{sec:append}, this implies that even
though the relative particle number-density increases with $\ed$,
the corresponding relative volume decreases in inverse proportion.

The scatter around {the FP} has a standard deviation $\sim 0.35$dex
and is typically larger than the corresponding observational errors
(mainly owing to distance measurement errors). This implies that the
scatter is due to some other systematic effects. Other unknown
parameters (i.e., $\alpha$, observer-angle, $\zeta$) may be
responsible for the thickening of the FP. We note that the
calculation of $L_{\gamma}$ in 2PC is based on the {observed flux},
$G_{\gamma}$, assuming that the {beaming-factor} $f_{\rm b}$ (see
\citealt{2010ApJ...714..810R}; 2PC) is 1 (i.e., the same) for all
the detected pulsars. However, our macroscopic and kinetic PIC
simulations show a variation of $f_{\rm b}$ with $\zeta$, which in
combination with the various $\alpha$-values could explain the
observed scatter. Therefore, the $L_{\gamma}$-values provided by
2PC, are essentially effective values, $L_{\rm \gamma\;eff}$, since
they are based on the assumption that the {corresponding $f_{\rm b}$
are} uniformly distributed.

The theoretical analysis, presented in this letter, provides a
simple physical justification of the observed FP based on the
assumption that $R_{\rm C}$ is a certain fraction/multiple of the
corresponding $R_{\rm LC}$, for all $\ed$. Nonetheless, the particle
orbits corresponding to different $\alpha$ and $\zeta$ values have
different $R_{\rm C}$ values. This implies that the proportionality
factor between $R_{\rm C}$ and $R_{\rm LC}$ varies with $\alpha$ and
$\zeta$, which consequently implies the existence of different
(though parallel) FPs. Thus, the relative position of a pulsar with
respect to the FP may constrain $\alpha$ and $\zeta$.

Any theoretical modeling should be able not only to reproduce the
uncovered relations but also to provide justifications of the
observed scatter. In a forthcoming paper, we will present under what
conditions kinetic PIC models reproduce the revealed $\gamma$-ray
pulsar sequence.

\acknowledgments We would like to thank an anonymous referee for
helpful suggestions that improved the letter. We also thank Ioannis
Contopoulos, Anatoly Spitkovsky, Isabelle Grenier, and David Smith
for stimulating discussions. This work is supported by the National
Science Foundation under Grant No. AST-1616632, by the NASA
Astrophysics Data Analysis Program, and by Fermi Guest Investigator
Program.

\appendix
\section{The Radius of Curvature Behavior in Arbitrary Electromagnetic Field Structure}
\label{sec:appendB}

In an electromagnetic field, an asymptotic trajectory is always
locally defined by the so-called Aristotelian electrodynamics
(\citealt{2012arXiv1205.3367G,2015AJ....149...33K}; K18)
\begin{equation}
\label{eq:ae}
\pmb{\mathrm{v_A}}=\frac{\mathbf{E}\times\mathbf{B}\pm(B_0\mathbf{B}+E_0\mathbf{E})}{B^2+E_0^2}c
\end{equation}
where $E_0B_0=\mathbf{E}\cdot\mathbf{B},\;\;E_0^2-B_0^2=E^2-B^2$.

%%%%%%%%%%%%%%%%%%%%%%%%%%%%%%%%%%%%%%%%%%%%%%%%%%%%%%%%%%%%%%%%%%%%%%%%
\begin{figure}[!hbt]
\vspace{0.0in}
  \begin{center}
    \includegraphics[width=0.8\linewidth]{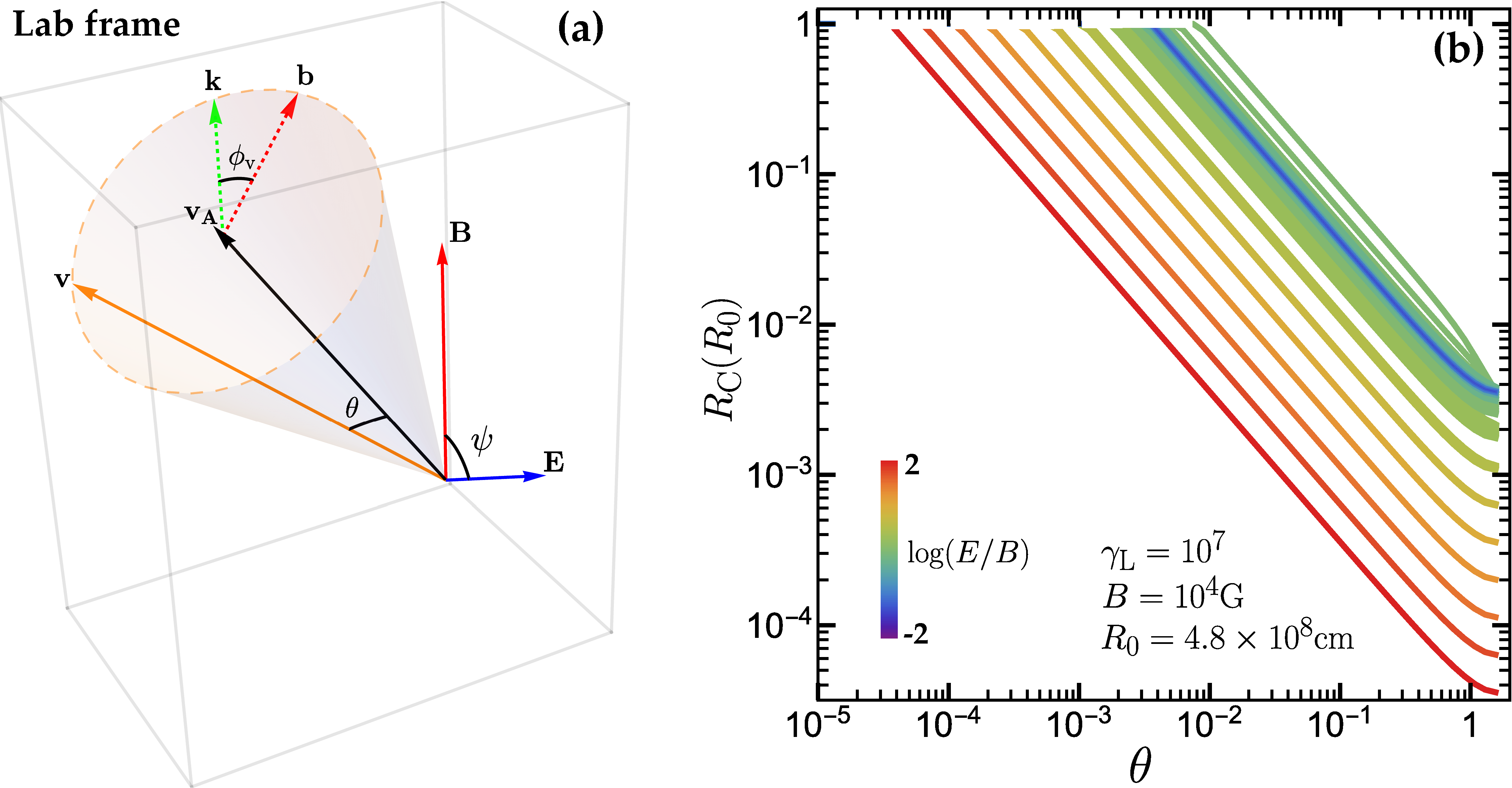}
  \end{center}
  \vspace{-0.1in}
  \caption{\textbf{(a)} The relative orientation between $\mathbf{v}$ and $\mathbf{v_A}$ is
  determined by the angles $\pa$ and $\phi_{\rm v}$ where $\mathbf{b}$, $\mathbf{k}$
  denote the perpendicular projection of $\mathbf{B}$ to $\mathbf{v_A}$ and
  the $\mathbf{v}\times\mathbf{v_A}$ direction, respectively. \textbf{(b)} The average
  (over $\phi_{\rm v}$) $R_{\rm C}$ for the indicated parameter values, $E/B$-ratio,
  and various $\psi$-values. For $E\ll B$ and $E\gg B$, the effect of $\psi$
  is negligible (thin single color regions) while for $E\approx B$, $\psi$ slightly modulates $R_{\rm C}$
  (green line zones).}
  \label{fig_app_B}
  \vspace{0.0in}
\end{figure}
%%%%%%%%%%%%%%%%%%%%%%%%%%%%%%%%%%%%%%%%%%%%%%%%%%%%%%%%%%%%%%%%%%%%%%%%

The particle velocity $\mathbf{v}$ continuously approaches
$\mathbf{v_A}$ (i.e., the generalized pitch-angle $\pa$ decreases).
The particle energy loss-rate is determined by the local $R_{\rm
C}={\gl m_{\rm e}c^2}/{(q_{e}B_{\rm eff})}$, where $\gl,~m_{\rm
e},~q_{\rm e}$ are the Lorentz factor, the mass, and the charge of
the particle, respectively, $c$ the speed-of-light, and $B_{\rm
eff}$ reads (C16)
\begin{equation}
\label{eq:Beff} B_{\rm
eff}=\sqrt{(\mathbf{E}+\mathbf{v}\times\mathbf{B}/c)^2-(\mathbf{v}\cdot\mathbf{E}/c)^2}~.
\end{equation}
Figure~\ref{fig_app_B}a shows that $R_{\rm C}$ depends on $E, B$,
the angles $\psi$, $\pa$, and the relative orientation of
$\mathbf{v}$ on the $\theta$-cone (i.e., $\phi_{\rm v}$). On the one
hand, the lowest $R_{\rm C}$-value, $r_{\min}$, which is achieved
for high $\pa$ is mainly determined by the order of magnitude of the
highest field value ($B_{\rm eff}=\max(E,B)$) while the variation of
$\psi$ and $\phi_{\rm v}$ produces a modulation around a mean value
(Fig.\ref{fig_app_B}b). On the other hand, for
$\mathbf{v}=\mathbf{v_{A}}$, $B_{\rm eff}=0$. Assuming that $R_{\rm
0}$ is the $R_{\rm C}$-value corresponding to the asymptotic flow, a
small velocity component perpendicular to $\mathbf{v_{A}}$ (i.e.,
small $\theta$) is developed that imposes $R_{\rm C}=R_{0}$. For
motion near the LC, the fields are $\sim B_{\rm LC}$ and therefore
$r_{\min}\sim r_{\rm g}$.

\section{Derivation of the Theoretical Fundamental Plane Relations}
\label{sec:append} The spin-down power for a dipole field reads
\begin{equation}\label{eq:app01}
    \ed\propto B_{\star}^2 P^{-4}~.
\end{equation}
Assuming
\begin{enumerate}[(i)]
 \item emission at the LC near the ECS (i.e., fields of the order of $B_{\rm LC}$)
 and taking into account that
\begin{equation}\label{eq:app05}
    B_{\rm LC}\propto B_{\star}R_{\rm LC}^{-3}\propto B_{\star}P^{-3}
\end{equation}
and $R_{\rm LC}\propto P$, we get
\begin{equation}\label{eq:app02}
R_{\rm C}\propto
\begin{cases} R_{\rm LC}\propto P&\text{CR-regime}\\
 r_{\rm g}\propto \gl P^{3}B_{\star}^{-1}&\text{SR-regime}
\end{cases}
\end{equation}
and then from Eq.\eqref{eq:ecut} and \eqref{eq:app02}, we get
\begin{equation}\label{eq:app03}
    \gl\propto
\begin{cases}
\ec^{1/3}P^{1/3} &\text{CR-regime}\\
\ec^{1/2}P^{3/2}B_{\star}^{-1/2}&\text{SR-regime}
\end{cases}
\end{equation}
\item a balance between acceleration and radiative losses
\begin{equation}\label{eq:app04}
    E_{\rm BLC}B_{\rm LC}\propto\gl^4R_{\rm C}^{-2}
\end{equation}
where $E_{\rm BLC}$ is the $\eacc$ in $B_{\rm LC}$ units. From
Eqs.\eqref{eq:app05}-\eqref{eq:app04}, we get
\begin{equation}\label{eq:app06}
    E_{\rm BLC}\propto
\begin{cases}\ec^{4/3}P^{7/3}B_{\star}^{-1}&\text{CR-regime}\\
\ec&\text{{SR-regime}}
\end{cases}
\end{equation}
and consequently the luminosity of one-particle reads
\begin{equation}\label{eq:app07}
    L_{\gamma 1}\propto E_{\rm BLC}B_{\rm LC}\propto
\begin{cases}
    \ec^{4/3}P^{-2/3}&\text{CR-regime}\\
    \ec B_{\star}P^{-3}&\text{{SR-regime}}
\end{cases}
\end{equation}
\item that the total $\gamma$-ray luminosity $L_{\gamma}$ scales with
the number of emitting particles in the dissipative region,
$N_d=n_{\rm GJ\mhyphen LC}\;V_d$, where $n_{\rm GJ\mhyphen LC}$ is
the Goldreich-Julian number-density at the LC, $n_{\rm GJ\mhyphen
LC}\propto n_{\rm GJ\star}R_{\rm LC}^{-3}\propto
B_{\star}P^{-1}R_{\rm LC}^{-3}$ where $n_{\rm GJ\star}$ is the
Goldreich-Julian number-density on the stellar surface and $V_d$ the
volume of the dissipative region, which we assume that $V_d\propto
R_{\rm LC}^{3}$. Thus, $N_d\propto n_{\rm GJ\star}\propto
B_{\star}P^{-1}$ and taking into account Eq.\eqref{eq:app01}, we get
\begin{equation}\label{eq:app08}
    L_{\gamma}\propto L_{\gamma1}B_{\star}P^{-1}\propto
\begin{cases}
    \ec^{4/3}B_{\star}^{1/6}\ed^{5/12}&\text{CR-regime}\\
    \ec\;\ed&\text{{SR-regime}}
\end{cases}
\end{equation}
\end{enumerate}

We note that according to Eqs.\eqref{eq:app01} and \eqref{eq:app08},
$L_{\gamma}$ may be a function of any 2-combinations of the
$(\ed,~B_{\star},~P)$ variable-set. Moreover, taking into account
that $\ed\propto P^{-3}\dot{P}$, $L_{\gamma}$ may also be expressed
as a function of the directly observable quantities
\begin{equation}\label{eq:app09}
    L_{\gamma}\propto
\begin{cases}
\ec^{4/3}P^{-7/6}\dot{P}^{1/2}&\text{CR-regime}\\
\ec\;P^{-3}\dot{P}&\text{{SR-regime}}
\end{cases}
\end{equation}
Nonetheless, any of these relations are equivalent.

\end{document}